\documentstyle[11pt,aaspp4]{article}

\slugcomment{To appear in The Astrophysical Journal, February 1, 1997}

\begin{document}

\def\lsim{\lower 4pt \hbox{$\, \buildrel {_{\textstyle 
<}}\over{\textstyle \sim}\,$}}

\def\deg{\ifmmode^\circ\else$^\circ$\fi}
 
\title{MULTIFREQUENCY RADIO OBSERVATIONS OF THE GRAVITATIONAL LENS SYSTEM
MG~0414+0534}
\author{C.~A. Katz, C.~B. Moore\footnote{Present address: Kapteyn Astronomical
Institute, Postbus 800, 9700 AV Groningen, The Netherlands}, \& J.~N. Hewitt}
\affil{Department of Physics and Research Laboratory of Electronics \\
Room 26-331,
Massachusetts Institute of Technology \\
 Cambridge, Massachusetts, 02139 \\
Electronic mail: ckatz@maggie.mit.edu, cmoore@maggie.mit.edu, jhewitt@maggie.mit.edu}

\begin{abstract}

The four-image gravitational lens system MG~0414+0534 was observed with
the VLA at 1.4, 5, 8, 15, and 22~GHz.  The 15 and 22~GHz images reveal
structure in the components which is consistent with that seen at other
wavelengths.  There was no detection of other extended emission, nor
were there detections of the lensing galaxy or the ``component~$x$''
seen in optical images.  Detections of polarized emission show that the
fractional polarized intensity is $\lsim 1$\%.  The observed properties
of MG0414 suggest that it is a GHz-peaked~spectrum source.

\end{abstract}


\section{Introduction}

Since the discovery of the ``double quasar'' 0957+561 by Walsh~{\it et
al.}~\markcite{w4}(1979), work on gravitational lensing has grown to
address many topics in astrophysics.  From observations of
multi-component systems and Einstein rings to theoretical studies of the
implications of lensing statistics on cosmology, the study of
gravitational lenses provides us with a unique probe of many phenomena.
Lenses may function as ``cosmic telescopes'' for studying background
sources; modeling lens gravitational potentials allows us to explore the
mass distribution in galaxies and clusters; measuring time delays
between the light curves for different images of the same source allows
us to estimate the values of cosmological parameters.  A review of some
applications of gravitational lenses is given by Blandford \&
Narayan~\markcite{b3}(1992).  To engage profitably in such pursuits
requires gravitational lens systems with well-measured characteristics.
MG~0414+0534 (also 4C05.19 and PKS0411+05; hereafter MG0414) is a
gravitational lens system that is very bright at radio wavelengths and
has optical and infrared counterparts.

MG0414 was discovered in the MIT-Green~Bank survey (Bennett {\it et
al.}~\markcite{b4}1986).  It was selected from the survey in a
gravitational lens search by Hewitt~\markcite{h3}(1986) and Hewitt~{\it
et al.}~\markcite{h4}(1989).  It was later studied in more detail by
Hewitt~{\it et al.}~\markcite{h5}(1992) and
Katz~\&~Hewitt~\markcite{k3}(1993).  The later observations showed that
MG0414 displays many of the characteristics expected in a gravitational
lens system: morphology typical of lensing by an elliptical potential,
similar spectral indices of the components, and frequency-independent
flux density ratios of the components.  While the optical flux ratios
reported by Schechter~\&~Moore~\markcite{s1}(1993),
Falco~\markcite{f3}(1993), and Angonin-Willaime~{\it et
al.}~\markcite{a1}(1994) differ significantly from those in the radio,
Witt, Mao, \&~Schechter~\markcite{w3}(1995) conclude from simulations of
microlensing in the MG0414 system that the differences between the radio
and optical flux ratios are consistent with microlensing models with
plausible physical parameters.  In addition, the observation of
significant reddening of the quasar images has fueled speculation that
the flux ratio differences are due at least in part to obscuration by
dust (Lawrence~{\it et al.}~\markcite{l3}1995; Vanderriest,
Angonin-Willaime, \& Rigaut~\markcite{v1}1996).

Other evidence in favor of gravitational lensing includes the discovery
of the lensing galaxy by Schechter~\&~Moore~\markcite{s1}(1993), the
measurement of similar optical spectra for the A and B components
(Angonin-Willaime~{\it et al.}~\markcite{a1}1994), and the measurement
of the source redshift of~2.639 (Lawrence~{\it et
al.}~\markcite{l3}1995).

Herein we describe VLA\footnote{The VLA is part of the National Radio
Astronomy Observatory, which is operated by Associated Universities,
Inc., under cooperative agreement with the National Science Foundation}
observations which were planned with several objectives in mind: {\it
high angular resolution}, to reveal structure in the components which
would be useful for constraining lens models; {\it sensitivity to
extended emission}, to map any extended structure, also useful in
modeling the lens and in examining the nature of the background source;
{\it measurement of polarization structure}, to provide information
about the background source and to measure Faraday rotation; {\it
multifrequency observations}, to measure the spectral indices of the
components; and {\it sensitivity to faint objects}, to search for a
fifth image and the lensing galaxy itself.

\section{Observations and Data Analysis}

\subsection{Observations}

MG0414 was observed with the A~configuration of the VLA at 15~GHz on
1992~November~18 and 1992~December~15, at 1.4, 5, 8, and~15 GHz on
1993~January~15, at 8 and 22~GHz on 1994~April~1, and with the
B~configuration at 15~GHz on 1993~March~9.  A summary of the
observations is given in Table~\ref{obssum}.
The 15~GHz runs in 1992 and on 1993~March~9 were part of a longer term
monitoring program, and were selected because they were the longest of
the monitoring runs that took place when the sky was clear.  During the
1993~January~15 observing, thick cloud cover and a snowstorm complicated
the calibration, especially at 15~GHz.  At that frequency, 200 minutes
were spent on MG0414, but about one-third of the data were discarded
because of very poor phase stability.  At the other frequencies, phase
stability was satisfactory.  Weather conditions during the 1994
observations were ideal; the VLA's reference pointing mode
(Perley~\markcite{p2}1995) was used at 22~GHz.  In all but the
monitoring observations the delay center was placed two arc~seconds east
of the center of the system, rather than directly at the center, to
avoid introducing any biases that might affect our search for a fifth
image or radio emission from the lensing galaxy.

\begin{deluxetable}{ccclc}
\tablewidth{450pt}
\tablecaption{Summary of Observations of MG 0414+0534\label{obssum}}
\tablehead{ 
& \colhead{Central}
\nl
\colhead{Date}          & \colhead{Frequency\tablenotemark{a}}
& \colhead{Integration} &
& \colhead{Map RMS}
\nl
\colhead{(UT)}          & \colhead{(GHz)}
& \colhead{Time}        & \colhead{Resolution\tablenotemark{b}}
& ($\mu$Jy/beam)
\nl
}
\startdata
92 Nov 18\tablenotemark{c} & 14.940 &$10^{\rm m}$ & $0.14'' \times 0.13''$, $-27^\circ$ &
400 \nl
92 Dec 15\tablenotemark{c} & 14.940 &$22^{\rm m}$ & $0.13'' \times 0.12''$, $+32^\circ$ &
280 \nl
93 Jan 15\tablenotemark{c} & 1.415 & $25^{\rm m}$ & $1.23'' \times 1.07''$, $-12^\circ$ &
610 \nl
& 4.860 &  $40^{\rm m}$  & $0.33'' \times 0.31''$, $-6^\circ$ & 200 \nl
& 8.440 &$40^{\rm m}$ & $0.24'' \times 0.19''$, $+6^\circ$ & 120 \nl
& 14.940 & $70^{\rm m}$
 & $0.11'' \times 0.10''$, $-23^\circ$ & 250 \nl
93 Mar 9\tablenotemark{d} & 14.940 & $20^{\rm m}$ & $0.69'' \times 0.37''$, $-46^\circ$ &
340 \nl
94 Apr 1\tablenotemark{c} & 8.440 & $5^{\rm m}$ & $0.24'' \times 0.22''$, $-2^\circ$
 & 130 \nl
 & 22.460 &  $95^{\rm m}$ & $0.08'' \times 0.08''$ & 330 \nl
\enddata
\tablenotetext{a}{At all frequencies $\Delta \nu = 100$~MHz, split into
two adjacent 50~MHz bands, except at 1.4~GHz, where the central
frequencies were 1365 and 1465~MHz.} 
\tablenotetext{b}{FWHM of beam major and minor axes and position angle
of major axis measured north through east}
\tablenotetext{c}{VLA A-configuration}
\tablenotetext{d}{VLA B-configuration}
\end{deluxetable}

\subsection{Calibration and Mapping}

In all data sets but one the flux density scales were set by reference
to 3C48.  In the 15~GHz data of 1993~January~15, the flux density scale
was set by reference to 3C286 because 3C48 was resolved on even the
shortest interferometric baselines.  Because 3C48 is resolved on all
baselines at 22~GHz, observations of it at that frequency were bracketed
by short scans on the phase calibrator 0133+476, allowing the creation
of a model of 3C48 which was used to set the flux density scale.

The VLA antenna gains at all frequencies were determined from short
observations of the calibrator source 0420-014.  This calibrator was
also used to determine the feed polarizations of the antennas.  In the
monitoring runs, 3C147 was used to fix the linear polarization position
angle.  All calibration and mapping was done with NRAO's Astronomical
Image Processing System (AIPS).  The maps shown here were produced after
several iterations of phase self-calibration, followed in some cases by
one iteration of amplitude self-calibration.  The self-calibration
significantly improved the dynamic range of the images.  Deconvolution
was performed with the AIPS task MX, restricting clean components to
small regions at the positions of the four objects.

\begin{figure}
\epsscale{0.9}
\plotone{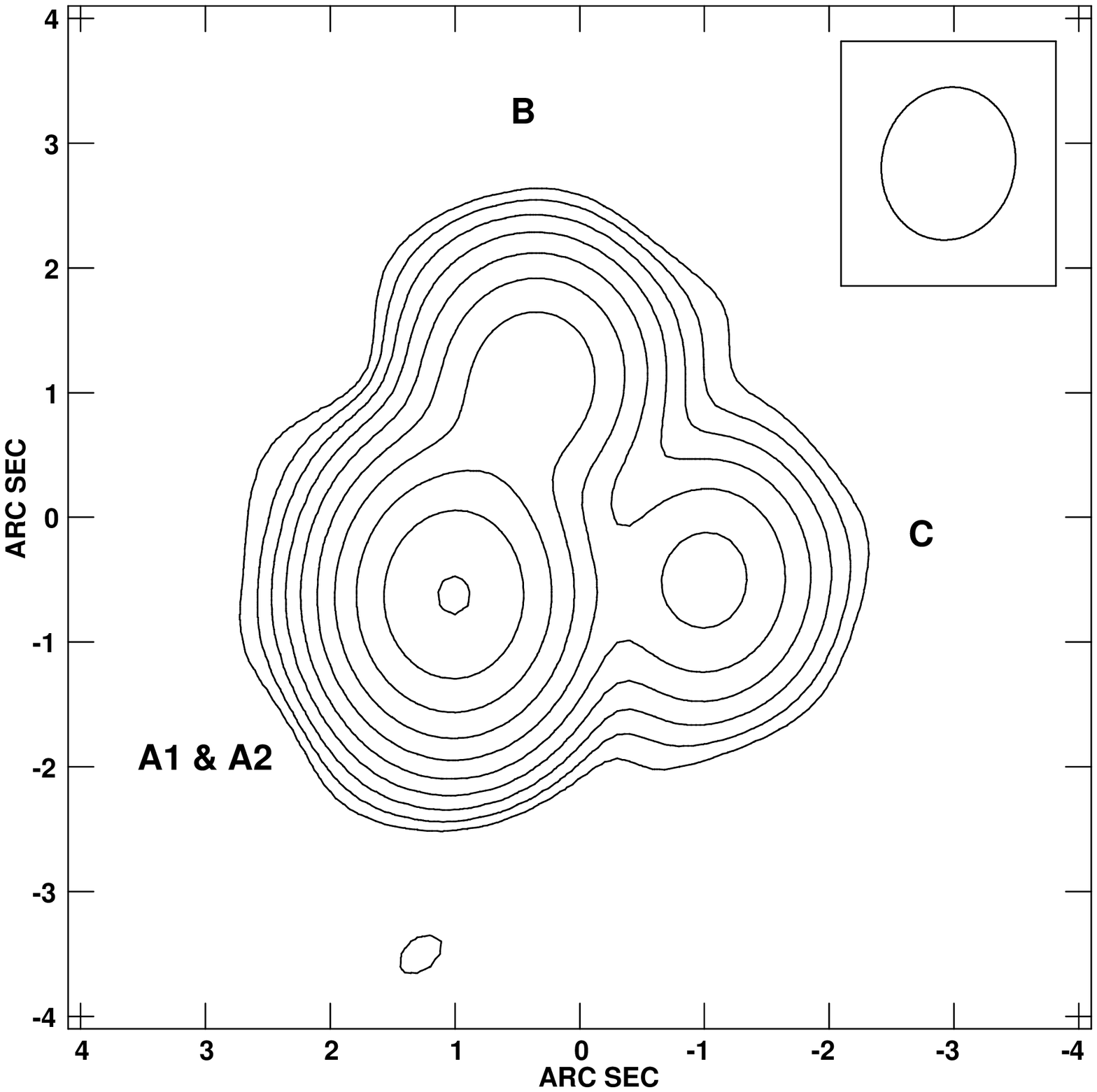}
\caption{Contour plot of 1.4~GHz map of MG~0414+0534 on 93~Jan~15.
  Contour levels are \mbox{-0.1875\%}, 0.1875\%, 0.375\%, 0.75\%, 1.5\%, 3\%,
  6\%, 12\%, 24\%, 48\%, and 96\% of the peak intensity of
  1.524~Jy/beam. The RMS noise level is 610~$\mu$Jy/beam.  The box in
  the upper right corner shows the beam FWHM ellipse.  The map is
  centered at RA~$04^h\,14^m\,37.708^s$,
  Dec~$+05^\circ\,\,34'\,\,43.10''$ (J2000).}
\label{lmap}
\end{figure}

\begin{figure}
\epsscale{0.9}
\plotone{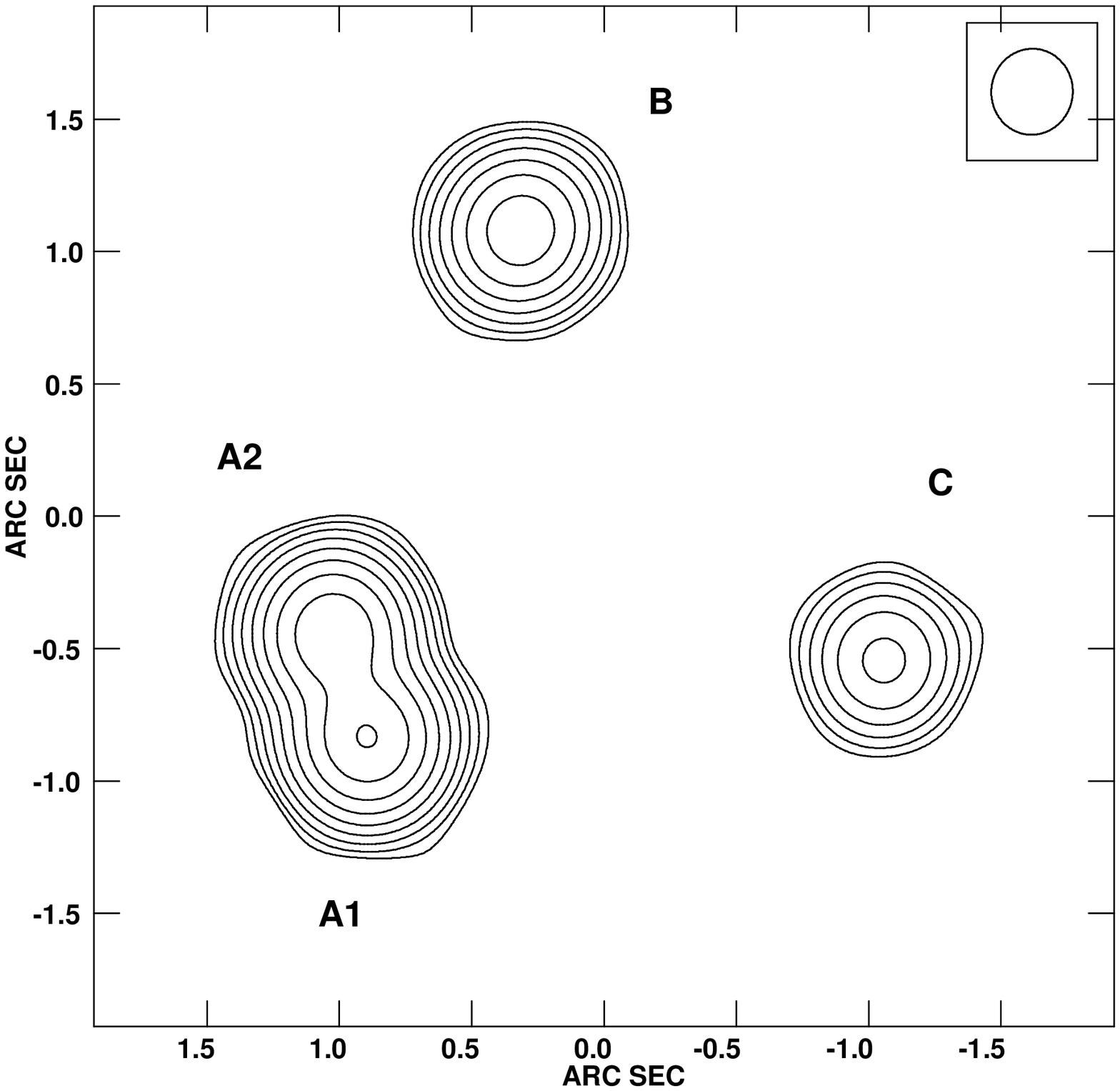}
\caption{Contour plot of 5~GHz map of MG~0414+0534 on 93~Jan~15.
  Contour levels are \mbox{-0.375\%}, 0.375\%, 0.75\%, 1.5\%, 3\%, 6\%, 12\%,
  24\%, 48\%, and 96\% of the peak intensity of 398~mJy/beam.  The RMS
  noise level is 200~$\mu$Jy/beam.  The box in the upper right corner
  shows the beam FWHM ellipse.  The map is centered at
  RA~$04^h\,14^m\,37.707^s$, Dec~$+05^\circ\,\,34'\,\,43.19''$ (J2000).}
\label{cmap}
\end{figure}

\begin{figure}
\epsscale{0.9}
\plotone{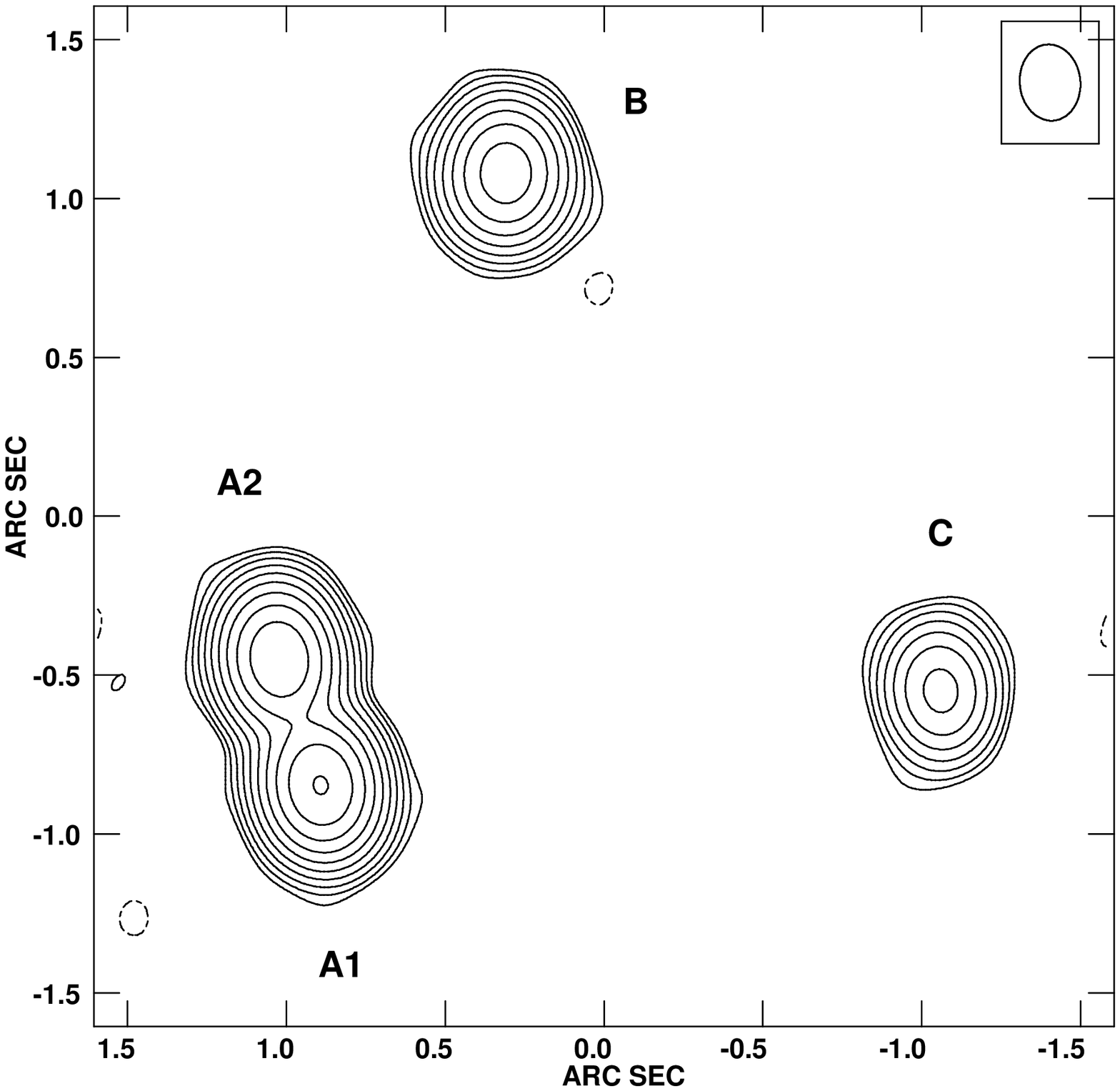}
\caption{Contour plot of 8~GHz map of MG~0414+0534 on 93~Jan~15.
  Contour levels are \mbox{-0.1875\%}, 0.1875\%, 0.375\%, 0.75\%, 1.5\%, 3\%,
  6\%, 12\%, 24\%, 48\%, and 96\% of the peak intensity of 238~mJy/beam.
  The RMS noise level is 120~$\mu$Jy/beam.  The box in the upper right
  corner shows the beam FWHM ellipse.  The map is centered at
  RA~$04^h\,14^m\,37.708^s$, Dec~$+05^\circ\,\,34'\,\,43.19''$ (J2000). }
\label{xmap}
\end{figure}

\begin{figure}
\epsscale{0.9}
\plotone{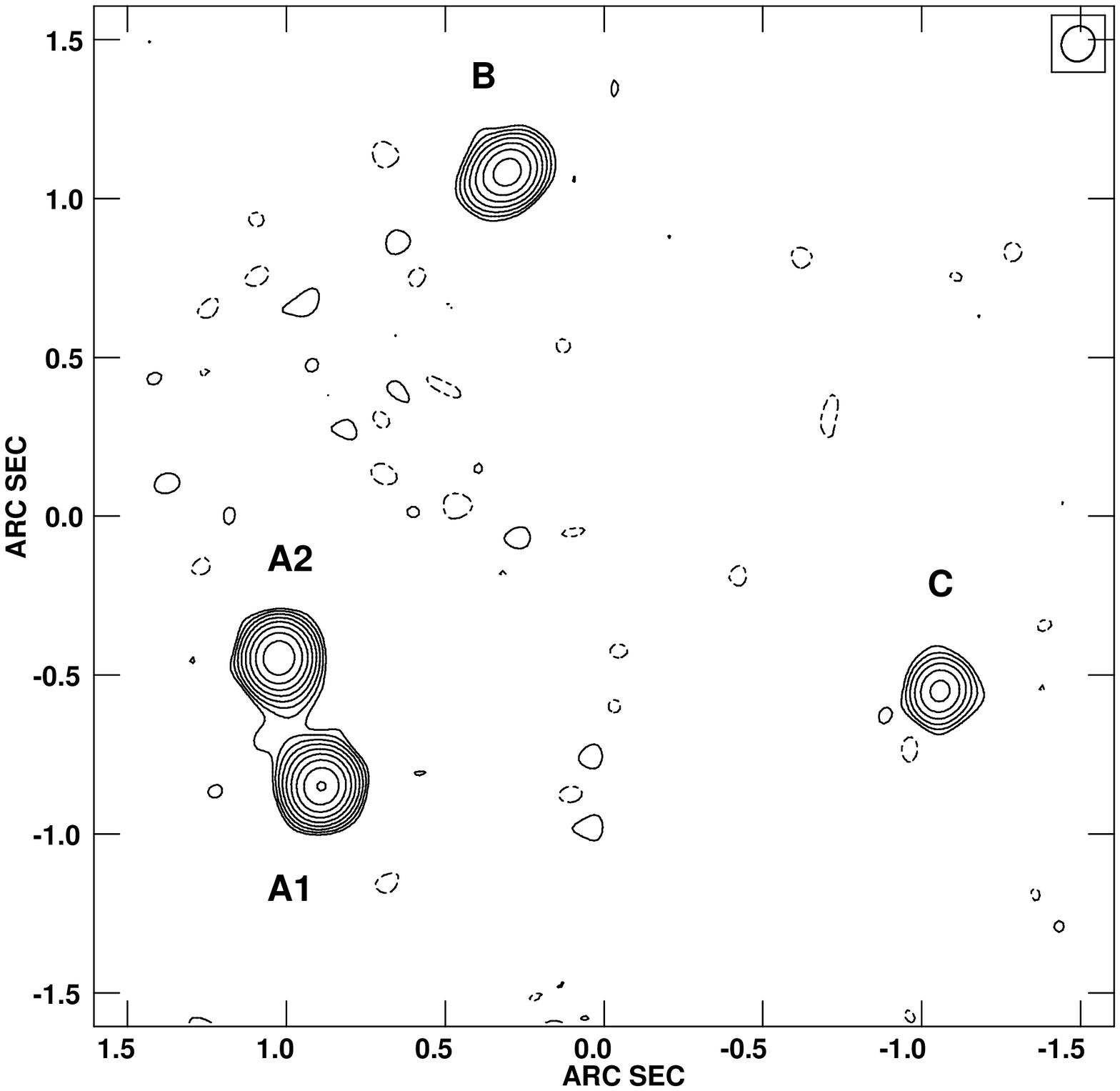}
\caption{Contour plot of 15~GHz map of MG~0414+0534 on 93~Jan~15.
  Contour levels are \mbox{-0.375\%}, 0.375\%, 0.75\%, 1.5\%, 3\%, 6\%, 12\%,
  24\%, 48\%, and 96\% of the peak intensity of 159~mJy/beam.  The RMS
  noise level is 250~$\mu$Jy/beam.  The box in the upper right corner
  shows the beam FWHM ellipse.  The map is centered at
  RA~$04^h\,14^m\,37.708^s$, Dec~$+05^\circ\,\,34'\,\,43.19''$ (J2000). }
  \label{umap}
\end{figure}

\begin{figure}
\epsscale{0.9}
\plotone{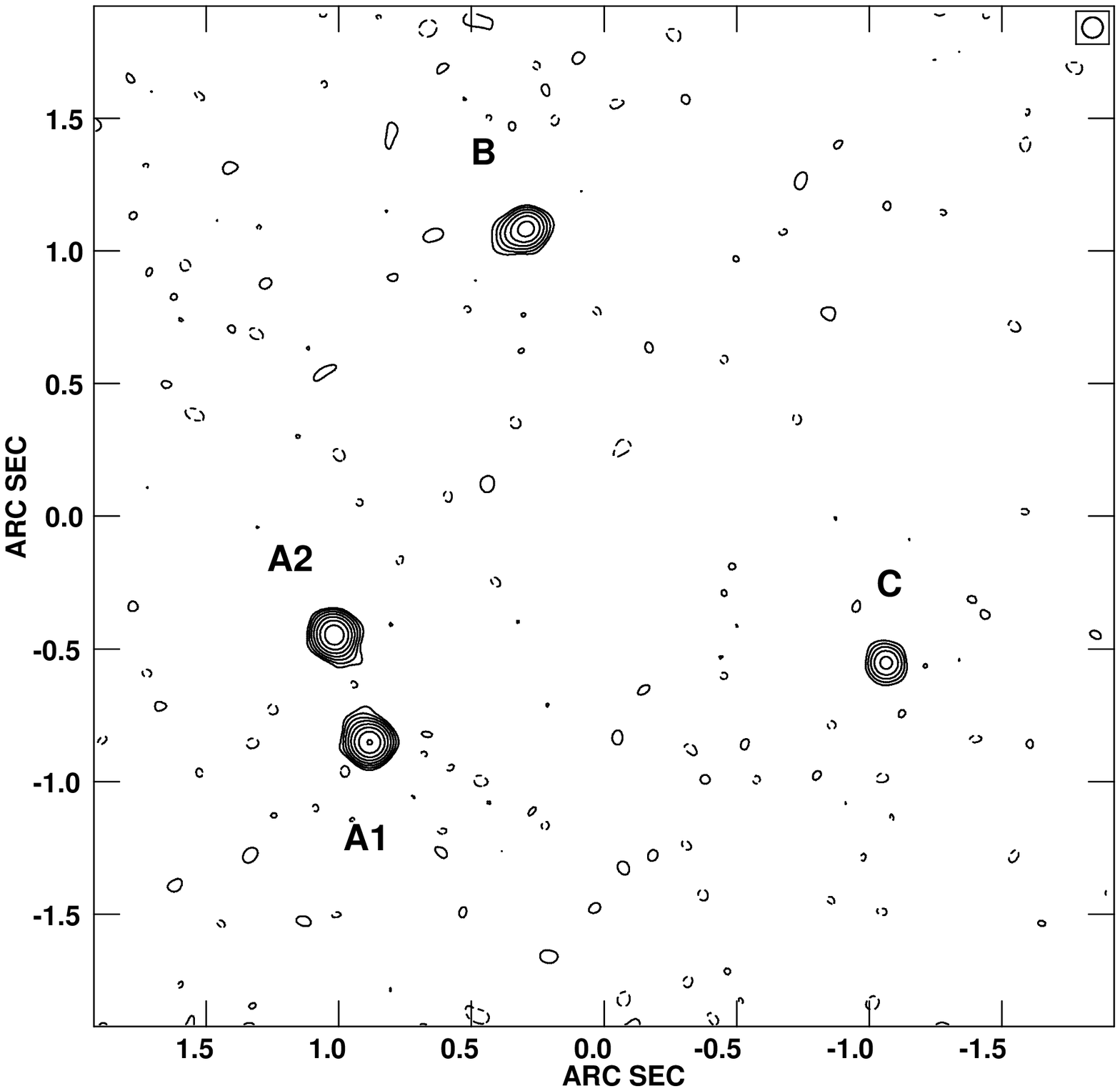}
\caption{Contour plot of 22~GHz map of MG~0414+0534 on 94~Apr~1.
  Contour levels are \mbox{-0.75\%}, 0.75\%, 1.5\%, 3\%, 6\%, 12\%, 24\%, 48\%,
  and 96\% of the peak intensity of 112~mJy/beam.  The RMS noise level
  is 330~$\mu$Jy/beam.  The box in the upper right corner shows the beam
  FWHM ellipse.  The map is centered at RA~$04^h\,14^m\,37.708^s$,
  Dec~$+05^\circ\,\,34'\,\,43.19''$ (J2000). }
\label{kmap}
\end{figure}

Figures \ref{lmap} through \ref{kmap} show, respectively, the 1.4~GHz
through 22~GHz maps produced from the 1993~January~15 observations. 
The dynamic ranges in the best maps are limited to about 2500:1 (peak
intensity:RMS noise level), which is much smaller than the dynamic range
of 10,000:1 expected if radiometer noise is the limiting factor.  Since
well-known methods for improving the dynamic ranges of VLA maps proved
unhelpful (see for example Perley~\markcite{p5}1989), we suspect that
the deconvolution procedure itself limited the dynamic range (see
Briggs~\& Cornwell~\markcite{b1}1994, Briggs~\markcite{b2}1995).  The
combination of the VLA dirty beam and the particular morphology of
MG0414 appears to exacerbate this effect; tests performed with noiseless
models of MG0414 confirm that this is indeed the case.  Further work in
investigating deconvolution algorithms is needed to improve the dynamic
ranges of VLA maps of MG0414.

The most notable feature in the high-resolution maps is the extended
structure: component~B is extended to the southeast, and components A1
and A2 are extended toward each other, with a faint connecting
``bridge'' of flux density approximately 2~mJy visible at 15~GHz.  The
extensions are consistent with the structure seen in a 5~GHz MERLIN
image (Garrett~{\it et al.}~\markcite{g1}1992), and with that seen in
the Hubble Space Telescope (HST) image of Falco, Leh\'ar, \&
Shapiro~\markcite{f4}(1996).  Our data provide upper limits on the
surface brightness of a fifth image of the background source and on any
emission associated with the lensing galaxy.  If we take these limits to
be 10 times the RMS surface brightness in each map, we find they are
6.1, 2.0, 1.2, 2.5, and 3.3~mJy at 1.4, 5, 8, 15 and 22~GHz,
respectively.  Expressed as a fraction of the flux density of
component~B the limits are 0.017, 0.013, 0.013, 0.038, and 0.072.  We do
not detect ``component~$x$,'' the sixth object west of component~B
detected at optical wavelengths by Schechter
\&~Moore~\markcite{s1}(1993), and confirmed by Angonin-Willaime~{\it et
al.}~\markcite{a1}(1994) and Falco~{\it et al.}~\markcite{f4}(1996).

The 1.4~GHz observation was intended to reveal any low surface
brightness extended emission which is likely to have a steep radio
spectrum and to give a $10\sigma$ detection of 70~mJy distributed
uniformly over a circle of radius $10^{\prime\prime}.$ An area the size
of the primary beam (about $31^\prime$ at 1.4~GHz) was mapped, but no
extended emission smaller than $38''$ (the maximum angular scale to
which this configuration of the VLA is sensitive) was found, even in
maps in which the short baselines were given large weights.

\subsection{Radiometry and Astrometry}

Absolute and relative radiometry of MG0414 (referred to component~B) are
presented in Tables~\ref{absrad}~\&~\ref{relrad}, respectively.
Except for the monitoring observations and the 8~GHz observation on
1994~April~1, the flux density uncertainties reported in
Table~\ref{absrad} were computed by comparing three effects: 1)~the
scatter in the amplitudes on unresolved baselines of the calibrated
uv-data for the primary flux calibrator, 2)~the uncertainty incurred in
the transfer of the antenna gain amplitudes from the primary flux
calibrator to the phase calibrator (0420-014), and 3)~the scatter in the
ensemble of flux density measurements made by splitting the dataset into
many short time segments and mapping each separately.  In all cases, 1)
was negligible with respect to 2).  Since 2) and 3) are independent, the
flux density uncertainties were found by adding them in quadrature.  For
the 15~GHz observation on 1993~January~15 which used 3C286 to set the
flux scale, an additional 2\% uncertainty was added in quadrature
reflecting the unknown systematic difference in the flux density scale
between 3C48 and 3C286.  The difference in elevation between the flux
calibration source and the phase calibrator was less than $6^\circ$ at
15 and 22~GHz, so elevation-dependent opacity differences should not
make a significant contribution to the error.  For the monitoring
observations, we estimate the error from the scatter in flux estimates
over the whole monitoring dataset (3.5\%)\footnote{Note that variability
in the source would make this an over-estimate of the error.}.  The
uncertainties in the relative flux densities reported in
Table~\ref{relrad} were computed from scatter of the A1/B, A2/B, and C/B
ratios in the segmented data described above.  The C/B ratio at 8~GHz is
significantly different from that reported in Katz
\&~Hewitt~\markcite{k3}(1993).  However, since the data used in Katz
\&~Hewitt~\markcite{k3}(1993) had a number of problems, and the ratios
in Table~\ref{relrad} are consistent with those measured from other
radio observations (e.g. Hewitt~{\it et al.}~\markcite{h5}1992), we
conclude that the difference in ratios cannot be regarded as reliable
evidence for variability.
\begin{deluxetable}{rrccccc}
\tablewidth{475pt}
\tablecaption{Absolute radiometry of MG 0414+0534.\label{absrad}}
\tablehead{  
                    & \colhead{Date of}
\nl
\colhead{Frequency} & \colhead{Observation}
& \colhead{A1}      & \colhead{A2}
& \colhead{B}       & \colhead{C}
& \colhead{Total}
\nl
}
\startdata
1.4 GHz & 93 Jan 15 & \multicolumn{2}{c}{$1644 \pm 8$\tablenotemark{a}}
	& $349 \pm 6$ & $125 \pm 6$ & $2118 \pm 11$ \nl
5 GHz & 93 Jan 15 & $401 \pm 8$ & $362 \pm 8$ & $156 \pm 4$ &
	 $58 \pm 2$ & $976 \pm 15$  \nl
8 GHz & 93 Jan 15 & $245 \pm 2$ & $220 \pm 2$ & $95 \pm 1$ & 
	$36 \pm 1$ & $596 \pm 4$  \nl
      & 94 Apr 1 & $242 \pm 1$ & $217 \pm 1$ & $93 \pm 1$ & 
	$37 \pm 1$ & $598 \pm 13$   \nl
15 GHz & 92 Nov 18 & $162 \pm 6$ & $142 \pm 6$ & $64 \pm 3$ &
	$23 \pm 1$ & $391 \pm 11$ \nl
       & 92 Dec 15 & $151 \pm 6$ & $134 \pm 5$ & $59 \pm 2$ &
	$23 \pm 1$ & $367 \pm 11$ \nl
       & 93 Jan 15 & $170 \pm 4$ & $150 \pm 4$ & 
	$66 \pm 2$ & $24 \pm 2$ & $411 \pm 10$  \nl
       & 93 Mar 9 & $145 \pm 6$ & $129 \pm 5$ & $56 \pm 2$ &
	$21 \pm 1$ & $351 \pm 10$ \nl
22 GHz & 94 Apr 1 & $117 \pm 6$ & $103 \pm 5$ & $46 \pm 3$ &
	$17 \pm 2$ & $282 \pm 11$  \nl
\enddata
\tablenotetext{}{Flux densities are in mJy referenced to 3C48.}
\tablenotetext{a}{Sum of the flux densities of A1 and A2.}
\end{deluxetable}
\begin{deluxetable}{rccccc}
\tablewidth{425pt}
\tablecaption{Relative radiometry of MG 0414+0534.\label{relrad}}
\tablehead{  & \colhead{Date of}
\nl
\colhead{Frequency} & \colhead{Observation}
& \colhead{A1} & \colhead{A2}
& \colhead{B} & \colhead{C}
\nl
}
\startdata
1.4 GHz & 93 Jan 15 & \multicolumn{2}{c}{$4.786 \pm 0.046$\tablenotemark{a}}
	& $1.000$ & $0.344 \pm 0.020$  \nl
5 GHz & 93 Jan 15 & $2.589 \pm 0.021$ & $2.325 \pm 0.016$ & $1.000$ &
	 $0.372 \pm 0.003$ \nl
8 GHz & 93 Jan 15 & $2.573 \pm 0.003$ & $2.309 \pm 0.003$ & $1.000$ & 
	$0.384 \pm 0.001$ \nl
15 GHz & 93 Jan 15 & $2.567 \pm 0.010$ & $2.286 \pm 0.009$ & 
	$1.000$ & $0.389 \pm 0.003$ \nl
22 GHz & 94 Apr 1 & $2.522 \pm 0.026$ & $2.221 \pm 0.022$ & $1.000$ &
	$0.396 \pm 0.006$ \nl
\enddata
\tablenotetext{a}{(A1~+~A2)/B.}
\end{deluxetable}

Astrometry of MG0414 is presented in Table~\ref{astrom}.
Since the measured positions were consistent at all five frequencies,
we show only the positions determined from the 22~GHz observations,
which had the highest resolution.  The position of component B was found
by fitting a two-dimensional gaussian.  The uncertainties in the
absolute position of B were taken to be $0.1''$, the highest accuracy
expected under normal observing conditions (Perley~\markcite{p2}1995).
The uncertainties in the positions of components A1, A2, and C relative
to B were calculated from the scatter in the segmented data (described
above).  The positions reported in Table~\ref{astrom} are consistent
with those reported by Falco~{\it et al.}~\markcite{f4}(1996) from HST
observations.
\begin{deluxetable}{ccc}
\tablewidth{350pt}
\tablecaption{22~GHz astrometry of MG 0414+0534.\label{astrom}}
\tablehead{  \colhead{Component} & \colhead{$\Delta\alpha$ ($''$)}
 & \colhead{$\Delta\delta$ ($''$)}
\nl
}
\startdata
A1 & $+0.5876'' \pm 0.0003''$    & $-1.9341'' \pm 0.0003''$ \nl
A2 & $+0.7208'' \pm 0.0003''$    & $-1.5298'' \pm 0.0003''$ \nl
B  & $\phantom{+}0.\phantom{7208'' \pm 0.0003''}$ & $\phantom{-}0.\phantom{5298'' \pm 0.0003''}$ \nl
C  & $-1.3608'' \pm 0.0007''$    & $-1.6348'' \pm 0.0008''$ \nl
\enddata
\tablenotetext{}{Positions are relative to component~B, which is located
at RA~$04^{\mbox{\small h}}$~$14^{\mbox{\small
m}}$~$37.7275^{\mbox{\small s}}$~$\pm$~$0.1''$, Dec~$+05^{\circ}$~$34'$~$44.276''$~$\pm$ $0.1''$ (J2000)}
\end{deluxetable}

\subsection{Polarimetry}

All data, except those taken at 15~GHz on 1993~January~15, were analyzed
for the polarization properties of the source.  Maps of the Stokes Q, U,
and V parameters were computed.  In the 1.4~GHz and 5~GHz data of
1993~January~15, there are nominal detections of polarized emission,
giving a polarization fraction of approximately 0.003 for A1 and A2.
However, the positions of the polarized emission peaks are significantly
different ($\sim\!0.15''$) from the positions of the total intensity
peaks, raising the question of the accuracy of the polarization
measurements.  The position offset we measure implies an uncertainty in
the fractional polarization of 0.006 (Roberts, Wardle, \&
Brown~\markcite{r1}1994); therefore, we conclude that the
1993~January~15 detections are spurious.  Since the observing conditions
on 1993~January~15 were less than optimal, and the absence of data from
an unpolarized source on that date precludes correcting for the
time-variable polarization response of the VLA, we derive information on
the polarization from the other datasets only.  Polarized emission was
detected from A1 and A2 at 15~GHz in the monitoring runs; for A1, the
measured polarization fraction is $0.0126 \pm 0.0011$ with position
angle $-46^\circ \pm 2^\circ$ (measured north through east).  For A2 the
polarization fraction is $0.0123 \pm 0.0009$ with position angle
$-52^\circ \pm 6^\circ$.  The errors are based on the scatter in the
data.  In the 8~GHz data of 1994~April~1 there were marginal detections
of polarized emission from A1 and A2 with polarization fractions less
than $0.01$, but the signal-to-noise ratios in the maps are so low that
we are unable to reliably estimate the polarization fractions or the
position angles.

\subsection{Spectral Indices}

\begin{figure}
\epsscale{0.9}
\plotone{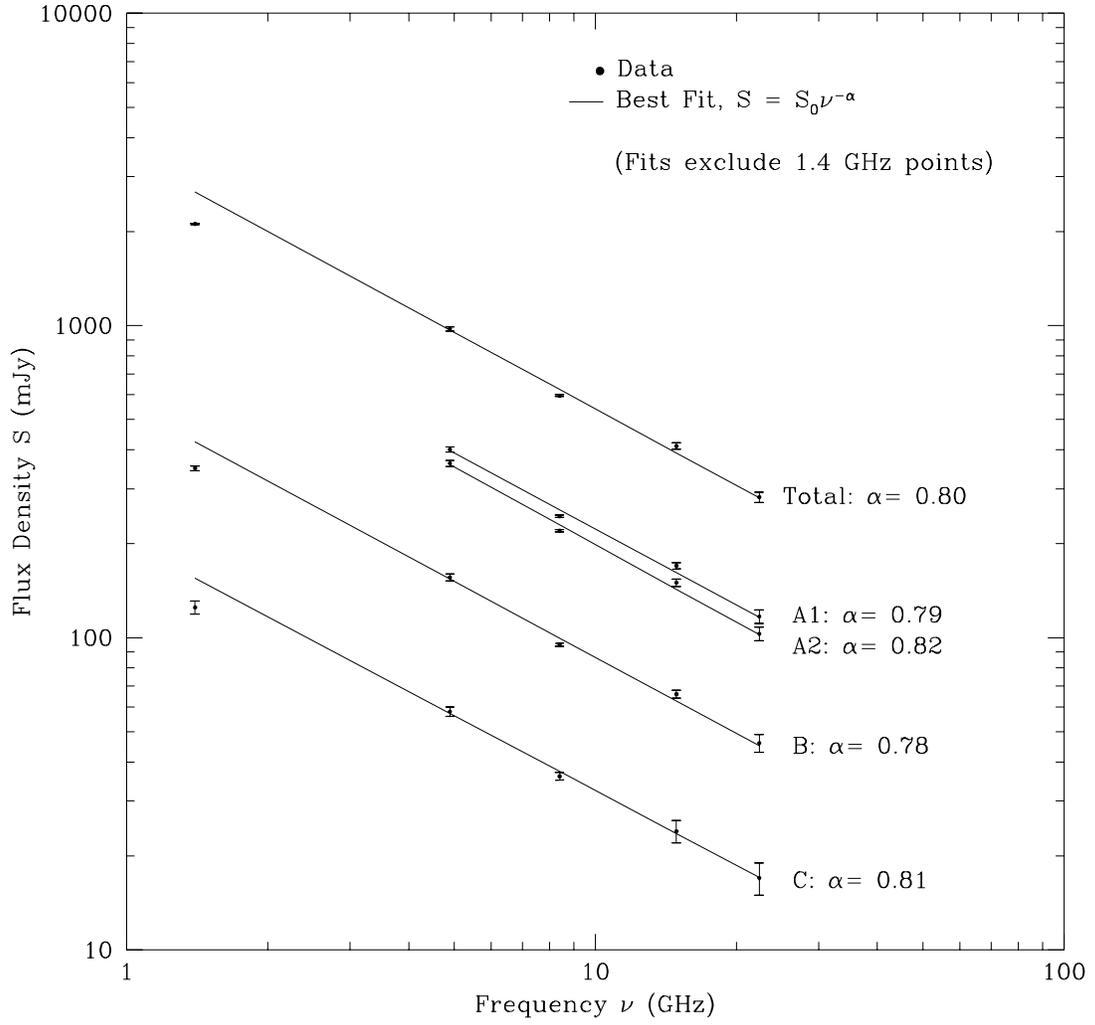}
\caption{Plot of measured flux densities of each component with spectral
model fits.  The fits exclude the 1.4~GHz data (see text).} \label{specplot}
\end{figure}

For each component, a spectral model of the form $S = S_o \nu^{-\alpha}$
was fit to the 5, 8, 15, and 22~GHz flux density measurements.  The 5,
8, and 15~GHz flux densities were taken from the simultaneous
measurements of 1993~January~15.  The results are shown in
Figure~\ref{specplot}.
Including the 1.4~GHz points in the fits made them unacceptable; we take
this as an indication that the spectral index $\alpha$ changes below
5~GHz.  The spectral indices for the four components are consistent;
averaging them yields an estimate for the source spectral index of
$\alpha = 0.80 \pm 0.02$.

\section{Discussion}

Quadruple-image lens systems are much better suited for gravitational
potential modeling than double-image lens systems because the additional
images provide more constraints.  However, even quadruple-image systems
do not provide unique lens models when only the position and flux
density of each component are measured (Gorenstein, Shapiro,
\&~Falco~\markcite{g3}1988; Kochanek~\markcite{k1}1991; Wambsganss
\&~Paczy\'nski~\markcite{w1}1994).  Resolving structure in the images is
required to constrain lens models.  The maps of MG0414 at 15 and 22~GHz
show structure on a scale comparable to the sizes of the synthesized
beams, making possible an improvement over the modeling effort of
Kochanek~\markcite{k1}(1991) and Hewitt~{\it et
al.}~\markcite{h5}(1992).  Modeling work is underway using both the VLA
data reported here and VLBI data (Ellithorpe, Hewitt,
\&~Kochanek~\markcite{e2}1996).  Our failure to detect ``component~$x$''
is not surprising in light of the fact that it is resolved in the HST
image, and thus appears to be a galaxy (Falco~{\it et
al.}~\markcite{f4}1996).

If the source is an isotropic radiator, the radio spectral luminosity
density implied by the observed flux density of component~B is $\sim
10^{29} h^{-2}$~erg~s$^{-1}$~Hz$^{-1}$ in the rest frame of the source
(assuming $H_0 = 100 h$~km~sec$^{-1}$~Mpc$^{-1}$, $q_0 = 0.5$, and
$\Lambda = 0$).  The magnification of the B image is not well
constrained by existing lens models (Hewitt~{\it et
al.}~\markcite{h5}1992; Kochanek~\markcite{k1}1991; Witt~{\it et
al.}~\markcite{w3}1995).  Values as large as 30 have been predicted,
which would reduce the inferred intrinsic luminosity by the same factor.
Even after correcting for the magnification, however, MG0414 has an
intrinsic luminosity comparable to those of powerful radio galaxies and
quasars.  Only a very small fraction of the radio emission detected with
the VLA is resolved, indicating a source size smaller than $1
h^{-1}$~kpc (assuming $H_0 = 100 h$~km~sec$^{-1}$~Mpc$^{-1}$, $q_0 =
0.5$, $\Lambda = 0$, and unit magnification).

At the low resolution of the single-dish survey of Condon, Broderick,
\&~Seielstad~\markcite{c4}(1991) the object appears as a double source;
the brighter of the two components is at the position of the MG0414
quasar images.  The fainter of the two is not seen in the 1.4~GHz data
in this work, implying a source size greater than $38''$. If the two
radio components were physically associated, their angular separation of
$12'$ would correspond to a linear distance of $2.7h^{-1}$~Mpc at the
source, at the high end of the distribution of radio source sizes
(Fanaroff \& Riley~\markcite{f2}1974).  However, there is no evidence in
the NRAO VLA Sky Survey (Condon {\it et al.}~\markcite{c5}1993) for
radio emission between the two sources, and it is unlikely that they are
physically associated.

Since the initial studies of MG0414 by Hewitt~{\it et
al.}~\markcite{h5}(1992), the nature of the background source has been
an issue because of its unusually red optical spectrum.  Observations in
several wavebands reported by Lawrence~{\it et al.}~\markcite{l3}(1995)
have led them to conclude that the source is a quasar made to look
unusual by the passage of the light rays through several magnitudes of
extinction due to dust in the lensing galaxy.  However, in the HST image
of Falco~{\it et al.}~\markcite{f4}(1996) there is an arc of emission
connecting component~B to the A1-A2 component pair.  The arc is
significantly less red than the image cores. Thus they argue that if the
arc and core sources have the same intrinsic color, and the arc-core
separation is small, then the reddening is intrinsic to the core source
rather than due to the lensing galaxy.

\begin{figure}
\epsscale{0.9}
\plotone{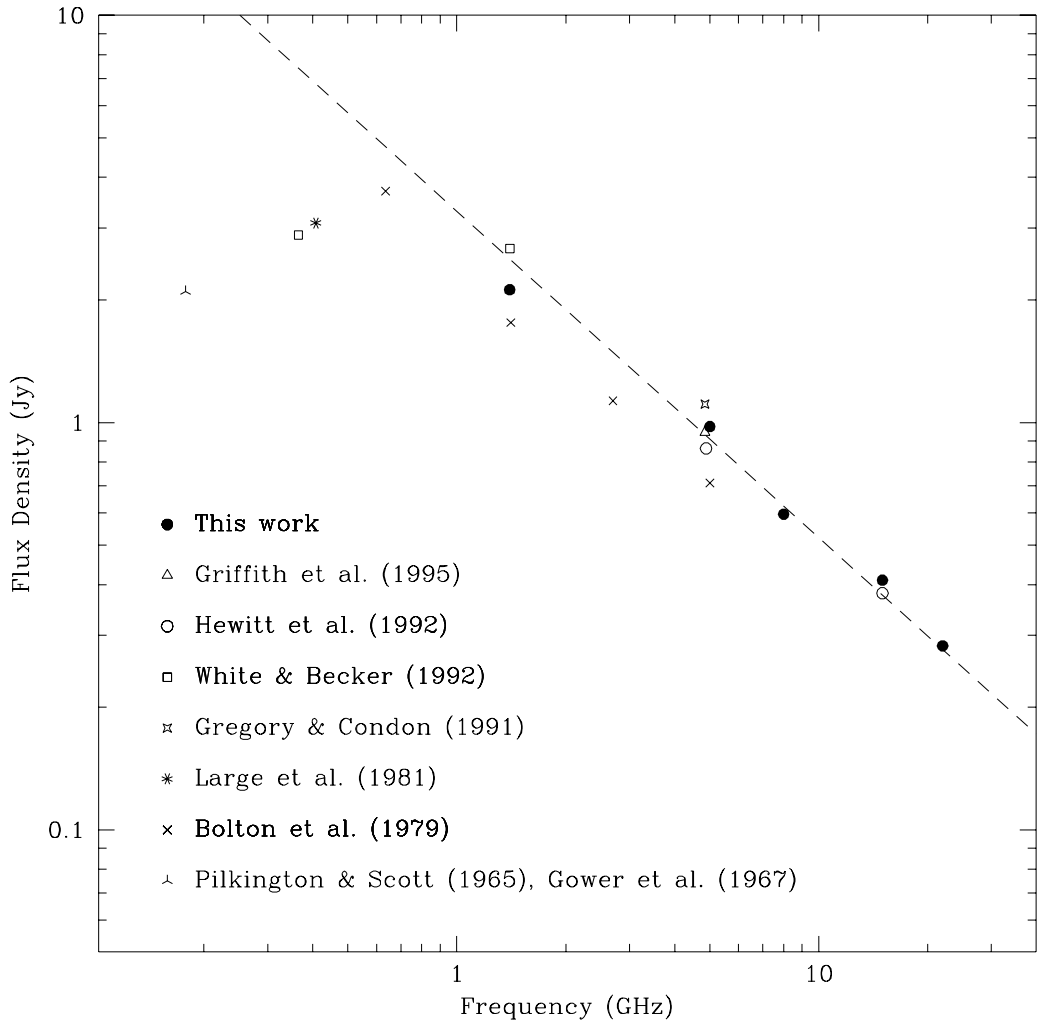}
\caption{MG~0414+0534 total flux density from this work and the
literature.  The dashed line indicates the power-law fit to data from
this work above 1.4~GHz (see text and Figure~\protect{\ref{specplot}}).}
\label{allflux}
\end{figure}

In Figure~\ref{allflux} we plot MG0414 flux density measurements from
this work and from the literature.
The radio spectrum turns over at low frequencies, characteristic of many
high redshift compact objects.  The small linear size and the peak of
the radio spectrum near 1~GHz suggest that MG0414 is an example of a
GHz-peaked spectrum (GPS) radio source.  The GPS sources form a class
closely related to, and perhaps a subclass of, the compact steep
spectrum (CSS) sources.  The GPS sources have sub-kiloparsec sizes and
spectral turnovers near 1~GHz, while the CSS sources have sizes of
1--10~kpc and spectral turnovers at one to several~hundred~MHz
(Fanti~{\it et al.}~\markcite{f1}1990, Kapahi~\markcite{k4}1981, Peacock
\& Wall~\markcite{p4}1982, Heckman~{\it et al.}~\markcite{h1}1994).  The
GPS sources are strongly correlated with high redshifts and large radio
luminosities (Gopal-Krishna, Patnaik, \&~Steppe~\markcite{g2}1983);
MG0414 has both properties.  The spectral index of MG0414 below the
turnover frequency is $\sim\!0.5$, intermediate to the ranges seen for
GPS and CSS sources.  Also, the low polarization fraction of MG0414 is
consistent with the properties of these classes.  Interestingly, another
gravitational lens, 0218+357, has also been identified as a GPS source
(O'Dea, Baum, \& Stanghellini~\markcite{o2}1991).

One possible explanation for the low polarization fraction is beam
depolarization, which occurs when the linear polarization angle varies
over a scale smaller than that of the synthesized beam.  Assuming $H_0 =
100 h$~km~s$^{-1}$~Mpc$^{-1}$, $q_0=0.5$, $\Lambda = 0$, unit
magnification, and a likely lensing galaxy redshift
of~0.5~(Kochanek~\markcite{k2}1992), the 5~GHz beamwidth of $0.3''$
corresponds to a linear distance of $1.05 h^{-1}$~kpc; at the source
redshift of $2.639$, $0.3''$ corresponds to a linear distance of $1.13
h^{-1}$~kpc.  Therefore, in order to reduce the polarization fraction by
50\%, the rotation measure must change by about 200~rad~m$^{-2}$ over a
distance of approximately 1.1~kpc.  This arises naturally in models of
GPS sources, but requires a lensing galaxy with unusual properties in
order to produce the required depolarization as well as the observed
optical reddening (Schechter~\markcite{s2}1995).

\section{Conclusions}

We have resolved some structure in components A1, A2, and B, at 15~and
22~GHz which is consistent with the arc seen in the HST image of
Falco~{\it et al.}~\markcite{f4}(1996).  We have not detected emission
from any other extended sources, from the lensing galaxy, or from
``component $x$.''  It is unlikely that the observed structure will
suffice to impose strong constraints on models of the radial mass
distribution of the lens.  Further work should address the
deconvolution-imposed dynamic range limit in the maps, to reach the
theoretical sensitivity limit and possibly resolve more low
surface-brightness structure.

The spectral indices of the components at frequencies above 1.4~GHz are
consistent with each other.  Averaging them yields a best estimate for
the source spectral index of $\alpha = 0.80 \pm 0.02$.

The fractional polarized flux density from MG0414 is very low, and is
detected at the threshold where our polarization calibration becomes
uncertain.  To exploit the polarization of the source for time delay
measurements or other studies, a program designed for more careful
calibration of the polarization properties of the VLA will be required.

MG 0414+0534 is a powerful high redshift quasar in which the effects of
obscuration in the source (or less likely, in the lens) are extremely
important. The source spectral peak at approximately 1~GHz, the compact
structure, the high redshift, the large radio luminosity, and the low
polarization fraction are all consistent with the GHz-peaked~spectrum
(GPS) sources.  GPS sources are thought to be AGN which are confined to
a very small space by an exceptionally dense circumnuclear gas.  If the
background source in MG0414 is indeed a GPS source, then the
gravitational lens may provide an unusual opportunity to study a member
of the GPS class through a ``cosmic telescope.''  We are presently
processing VLBA data which will address this possibility.

\acknowledgments

We wish to thank Rick Perley for his advice on VLA observing techniques,
and Pauline~McMahon for her help with the actual observations.  Thanks
are also due Jim Condon for providing us with an early map from the NRAO
VLA Sky Survey.  We are grateful to Emilio Falco for providing HST maps
of MG0414, and for his useful suggestions.  Finally, we thank the
anonymous referee for helpful comments.  This work was supported by a
David and Lucile Packard Fellowship in Science and Engineering, a
National Science Foundation Presidential Young Investigator Award, and
the M.I.T. Class of 1948.

%

%

\end{document}